\documentclass[11pt]{article}
\usepackage{jheppub}
\usepackage{amsmath,amssymb,amsfonts,graphicx,slashed,color, mathtools, amsthm}
\usepackage{comment}
\usepackage{enumerate}
\usepackage{float}
\usepackage{caption}
\usepackage{subcaption}
\usepackage{pdflscape}
\usepackage{booktabs}

\newcommand{\be}{\begin{equation}}
\newcommand{\ee}{\end{equation}}
\newcommand{\bea}{\begin{eqnarray}}
\newcommand{\eea}{\end{eqnarray}}
\newcommand{\beas}{\begin{eqnarray*}}
\newcommand{\eeas}{\end{eqnarray*}}
\newcommand{\ba}{\begin{array}}
\newcommand{\ea}{\end{array}}

\title{Suggestions of decreasing dark energy from supernova and BAO data: an update}

\author[1]{Mark Van Raamsdonk,}
\emailAdd{mav@phas.ubc.ca}

\author[2]{Chris Waddell}
\emailAdd{cwaddell@perimeterinstitute.ca}

\affiliation[1]{Department of Physics and Astronomy, University of British Columbia,\\
6224 Agricultural Road, Vancouver, B.C.\ V6T 1Z1, Canada}
\affiliation[2]{Perimeter Institute for Theoretical Physics, 31 Caroline St N, Waterloo, ON\ N2L 2Y5, Canada.}

\abstract{In a previous work 2305.04946, we found that supernova and baryon acoustic oscillation data support the hypothesis that late time cosmic acceleration is caused by the potential energy of a scalar field descending its potential, as suggested by holographically defined models of quantum gravity. In this note, we update our analysis using the Dark Energy Survey 5 year supernova data set (DES-SN5YR) and the baryon acoustic oscillation data from the Dark Energy Spectroscopic Instrument Data Release 2 (DESI DR2). Approximating the scalar potential via a first order Taylor series $V \approx V_0 + V_1 \phi$ about the present value, and making use of only recent-time data from DES-SN5YR and DESI DR2, we find that the slope parameter is constrained as $V_1 = 1.49 \pm 0.25$ in a standard likelihood analysis. This is naively a $>5 \sigma$ discrepancy with $\Lambda$CDM (which has $V_1 =0$), though a more detailed analysis not assuming a Gaussian likelihood distribution suggests $4 \sigma$ significance. Based only on the $\Delta \chi^2 = -13.7$ improvement of fit while ignoring parameter space volumes disfavours $\Lambda$CDM at a $3 \sigma$ significance level. These significance measures are substantially improved from our previous analysis using older data sets. We also reproduce the DESI DR2 parameter constraints based on the same combination of data and find that the $\Lambda$CDM is more strongly disfavoured in the context of the linear potential extension (dubbed $V_0V_1$) as compared with the $w_0 w_a$ extension of $\Lambda$CDM.  A caveat is that for both $w_0 w_a$ and $V_0 V_1$, much of the significance relies on the historical $z < 0.1$ supernova samples included in the DES-SN5YR data set.
}
\keywords{}

\begin{document}

\maketitle

\section{Introduction}

A particularly ironic twist in the recent history of science is that the 1997 discovery of the AdS/CFT correspondence (a.k.a. holography) in string theory was followed almost immediately by the 1999 discovery of the accelerated expansion of the universe in late time cosmology. The former provided, for the first time, fully defined UV-complete theories of quantum gravity, but with the restriction that these gravitational theories have a negative cosmological constant. The simplest explanation for the latter observations also requires introducing a cosmological constant, but this time, with a positive value. So the set of gravitational models known to be well-defined theoretically appeared to be the complement of the set of gravitational models that would be useful to explain observations of our universe.

During the past 25 years, there has been a substantial effort in the string theory community to extend holographic methods to define gravitational theories with positive cosmological constant, but progress has been surprisingly difficult \cite{Banks:2001px,Strominger:2001pn,Alishahiha:2004md,Gorbenko:2018oov,Coleman:2021nor,Freivogel2005,McFadden:2009fg,Banerjee:2018qey,Susskind:2021dfc,Silverstein:2024xnr} and there are even arguments that that it may not be possible \cite{Obied:2018sgi}. 

In \cite{VanRaamsdonk:2022rts} (see also \cite{Antonini:2022blk}), we have argued that the generalization to positive cosmological constant may not be necessary. We emphasized that generic cosmological solutions of holographically defined gravitational theories have scalar fields that vary on cosmological timescales. The potential for these scalar fields has an extremum with negative value - this value is the negative cosmological constant - but away from this extremum, the potential typically rises to positive values. In a cosmological solution, the scalar field can start at a positive value of its potential and roll down to a negative value, leading to a period of accelerated expansion before eventual deceleration and recollapse.

Thus, despite their negative cosmological constant, the quantum gravity theories that we know how to define holographically have generic cosmological solutions with time-varying scalars that can potentially explain the observed accelerated expansion.\footnote{At the level of effective field theory, the mechanism is simply a quintessence model for dark energy \cite{Peebles:1987ek,Ratra:1987rm,Caldwell:1997ii,Dutta:2018vmq,Visinelli:2019qqu,sen2021cosmological}.} These holographic considerations come with a prediction \cite{Antonini:2022blk,VanRaamsdonk:2022rts}: that the scalar potential energy should typically evolve by an order one amount on a Hubble timescale as the scalar descends its potential towards the negative extremum. In other words, dark energy should be dynamical, and it should not be a small effect. 

Based on these expectations (and 
prior to the recent DESI results \cite{DESI:2024mwx} suggesting dynamical dark energy), we sought to test these expectations by comparing with observational data in \cite{VanRaamsdonk:2023ion}, finding that supernova (SN) and baryon acoustic oscillation (BAO) data provides significant support for a model with decreasing dark energy via a time-dependent scalar (details below). Since that analysis, newer higher-quality supernova and BAO datasets have become available. In this paper, we repeat our earlier analysis using the new datasets and find substantially increased significance. 

\subsubsection*{Methods}

We now review briefly the methods used in our earlier analysis \cite{VanRaamsdonk:2023ion} and repeated here.

The holographic considerations do not suggest any specific form for the scalar potential except that it should have an extremum with negative $V$ and interaction terms that lead the potential to become positive away from the extremum. Given our ignorance of the specific form of the potential, we parameterize the scalar potential via a Taylor series about its present value,\footnote{Here we are taking units where $8 \pi G / 3 = H_0 = 1$ so $V_0 = \Omega_\Lambda$ in a $\Lambda$CDM model. With our normalization of $V_1$, an order one value of $V_1$ results in a order one fractional change in the scalar potential over a Hubble time.}
\[
V(\phi) = V_0 + V_1 \phi + V_2 \phi^2 + \ldots
\]
For times that are relatively small compared to a Hubble time, we expect that keeping only the linear terms should be a good approximation. Thus, the model we consider is equivalent to a linear potential quintessence model.\footnote{ Analyses of such a model based on older data sets may be found in \cite{Kallosh:2003bq, wang2004current, Perivolaropoulos:2004yr, Avelino:2004vy, Sahlen:2006dn, Huterer:2006mv}, but the previous data was not constraining enough to show any preference of the linear potential model over $\Lambda$CDM. See also \cite{Dimopoulos:2003iy, Efstathiou:2008ed} for other early appearances of the linear potential model. } Crucially, we don't necessarily expect this form of the potential to be valid over the full period of cosmological evolution, so we constrain parameters only with late-time data, specifically type Ia supernovae and baryon acoustic oscillations but not the cosmic microwave background.

In \cite{VanRaamsdonk:2023ion}, we made use of the redshift versus brightness data in the Pantheon+ \cite{Scolnic:2021amr} catalogue of type Ia supernovae, as well as BAO data from various surveys \cite{Ross:2014qpa, BOSS:2016wmc, eBOSS:2020lta, eBOSS:2020hur, eBOSS:2020fvk, eBOSS:2020uxp, eBOSS:2020gbb, eBOSS:2020tmo}. We found that a standard likelihood analysis based on this data significantly preferred models with $V_1 > 0$ in conventions where $\dot{\phi}^{\text{now}} \le 0$, with the $\exp(-\chi^2/2)$ distribution giving $V_1 = 1.1 \pm 0.3$. 

In this paper, we update the analysis with newer data sets that have become available since the previous work. For supernovae, we use the Dark Energy Survey five year supernova dataset (DES-SN5YR) \cite{DES:2024upw, DES:2024hip, DES:2024jxu}, while for baryon acoustic oscillations, we use the second data release (DR2) of the Dark Energy Spectroscopic Instrument \cite{DESI:2025zpo, DESI:2025zgx}. 

We find results that are consistent with the earlier analysis, but with substantially improved significance. Specifically, a Bayesian likelihood analysis finds parameter constraints\footnote{Throughout this note, we assume spatially flat cosmologies.} 
\[
\Omega_M = 0.292 \pm 0.021 \qquad V_0 = 0.627 \pm 0.022 \qquad  V_1 = 1.489 \pm 0.250 \; ,
\]
so the $\Lambda$CDM value of $V_1 = 0$ lies more than 5$\sigma$ away from the mean. 

To further assess the significance of our result, we generated an ensemble of mock datasets assuming a $\Lambda$CDM cosmology with the value $\Omega_{\rm M} = 0.310$ that best fits the SN and BAO data, adding random Gaussian noise drawn from the observational covariance matrices. We analyzed each using the $V_0 V_1$ linear potential model. The resulting distribution of best-fit values exhibited a bias toward positive $V_1$, with a mean of 0.73 and a standard deviation of 0.19 in the distribution of mean $V_1$ values over the ensemble. This positive $V_1$ bias arises because $V_1 < 0$ values give behavior that deviates much more strongly from $\Lambda$CDM behavior than comparable magnitude $V_1 > 0$ values. Nevertheless, the value inferred from the real data, $V_1 = 1.49$, lies well outside the distribution of average $V_1$ values produced by mock $\Lambda$CDM realizations, suggesting $4 \sigma$ significance (calculated as  $4 = (1.49 - 0.73)/0.19$).

Comparing $\chi^2$ for the best fit linear potential model to that of the best fit $\Lambda$CDM model, we find $\Delta \chi^2 \approx -13.7$, associated with a $3\sigma$ significance level since we are working with a two-parameter extension of $\Lambda$CDM where $V_1 = 0$ and $V_0 = 1 - \Omega_M$. The lower significance value obtained with this method arises because a relatively tiny volume of parameter space near the best fit $\Lambda$CDM model has a comparable $\chi^2$ to that model, while a much larger volume of parameter space near the best fit linear potential model has comparable $\chi^2$ to the minimum for that model.

It is interesting to compare our results with results of the DESI collaboration \cite{DESI:2024mwx,DESI:2025zgx} for the $w_0 w_a$ model that have recently received a lot of attention. As a check of our methods, we repeated the analysis of parameter constraints on the $w_0w_a$CDM models reported in \cite{DESI:2025zgx}, reproducing the parameter constraints reported there based on the DES-SN5YR and DESI DR2 data. We note that for the $w_0 w_a$ model compared to $\Lambda$CDM, $\Delta \chi^2 \approx -13.5$, so the linear potential model allows a slightly better fit to the data. The difference is more significant when looking at the $\exp(-\chi^2/2)$ distributions (Figure \ref{fig:V0V1w0wa}) where $\Lambda$CDM lies substantially further outside the 68\% and 95\% confidence regions in the $V_0 V_1$ case. 

While both $V_0 V_1$ and $w_0 w_a$ represent two-parameter extensions of $\Lambda$CDM at the level of the Friedmann equations, the scalar field model has the advantage that it is based on a simple physical effective field theory. In particular, it automatically satisfies $w \ge -1$ for the full evolution. In contrast, the $w_0 w_a$ models found in \cite{DESI:2024mwx,desicollaboration2025desidr2resultsii} to improve upon $\Lambda$CDM typically have $w < -1$ for a significant part of their evolution. This behavior is associated with a violation of the null energy condition and may be difficult to achieve in a sensible physical theory. In the $V_0 V_1$ models, $w(a)$ is significantly non-linear \cite{VanRaamsdonk:2023ion}, so is not represented well by any $w_0 w_a$ model.

We find that extending the model to include a quadratic term in the potential does not significantly improve the fit and that removing data at high or low redshift broadens the $\exp(-\chi^2/2)$ distribution in a way that remains consistent with the results from the full data set. A caveat is that the significance of the preference of both $V_0 V_1$ and $w_0 w_a$ models over $\Lambda$CDM drops considerably when excluding a set of historical $z < 0.1$ supernova observations included in the DES-SN5YR data set (see Section \ref{sec:red_data_sets}). This could indicate that the suggestions of decreasing dark energy originate from unaccounted for systematic errors in this older data for nearby supernovae. The importance of low $z$ data for the significance of the $w_0w_a$ results was also pointed out in \cite{DESI:2025zgx,Gialamas:2024lyw,Efstathiou:2024xcq,Notari:2024zmi,Colgain:2024mtg,Huang:2025som}. 

Simple scalar potential models including the linear potential have also been compared with supernova and BAO data recently in \cite{Berghaus:2024kra}, finding a preference for quintessence models over $\Lambda$CDM.\footnote{See \cite{Bhattacharya:2024hep, Shlivko:2024llw, Wolf:2024eph, Bhattacharya:2024kxp,Payeur:2024dnq,Akthar:2024tua,Hossain:2025grx} for other recent discussions of implications of new cosmological data on quintessence models.} A key difference is that in our work, we view the linear form of the potential as a Taylor approximation that is likely to be valid only in the relatively recent cosmological past, so we do not try to incorporate CMB data when constraining parameters. In this sense, our results are complementary to those of \cite{Berghaus:2024kra} who make use of both CMB and SN/BAO data (employing DESI DR1 for BAO).

In the remainder of the paper, we review our methods in Section \ref{sec:methods}, provide detailed results in Section \ref{sec:results}, and give a brief discussion in Section \ref{sec:disc}.

\section{Methods} \label{sec:methods}

The theoretical background and methods used in the present analysis are the same as in \cite{VanRaamsdonk:2023ion} so we only briefly review them here and refer to that work for more details.

\subsubsection*{Model}

Throughout our analysis, we consider a spatially flat FRW universe, with metric
\be
ds^2 = -dt^2 + a(t)^2 d \vec{x}^2 \: ,
\ee
and take $a=1$ presently. We consider a single scalar field  $\phi$ minimally coupled to gravity with potential $V(\phi)$. The evolution equations for the metric and scalar field are then
\be
\label{Friedmann}
H^2 = {8 \pi G \over 3}\left[\tilde{\rho} + {1 \over 2} \dot{\phi}^2 + V(\phi) \right]
\ee
and
\be
\label{scalarEOM}
\ddot{\phi} + 3 H \dot{\phi} +  V'(\phi) = 0 \: ,
\ee
where a dot denotes a $t$ derivative, $H = \dot a/a$ is the Hubble parameter, and $\tilde{\rho}$ is the energy density in all species excluding that of the scalar field. 

We will only consider cosmological evolution on recent timescales $z<2.33$ where the effect of radiation is negligible, so we have to a very good approximation that $\tilde{\rho}$ is dominated by matter,
\begin{equation}
\tilde{\rho} = {\rho_{M}^{\text{now}} \over a^3} = {3 H_0^2 \Omega_M \over 8 \pi G} {1 \over a^3}  \; ,
\end{equation}
using the usual definition of $\Omega_M$, with $H_0$ defined to be the present value of $\dot{a}/a$. 

We can define a dimensionless field
\begin{equation}
\hat{\phi} = \sqrt{8 \pi G \over 3} \phi
\end{equation}
and write the potential as
\begin{equation}
V(\phi) = {3 H_0^2 \over 8 \pi G} \left[V_0 + V_1 \hat{\phi} + {1 \over 2} V_2 \hat{\phi}^2 + \cdots\right] \: ,
\end{equation}
where the parameters in the potential are dimensionless.

Finally we can define a dimensionless time parameter $s = t H_0$. In this case, the equations simplify to 
\begin{equation}
\label{Friedmann}
\begin{split}
    \hat{H}^2 &= {\Omega_M \over a^3} + {1 \over 2} \dot{\phi}^2 + V(\phi) \\
    0 & = \ddot{\phi} + 3 \hat{H} \dot{\phi} +  V'(\phi)
\end{split}
\end{equation}
where $\hat{H} \equiv a^{-1} da/ds$ and we omit the hats on $\phi$ here and below.\footnote{The dimensionless quantity $\hat{H}$ is often referred to as $E$ elsewhere in the literature.}

In our analysis, it will be convenient to define $v = d \phi / ds$ and define a conformal time $\eta$ via $d \eta = -dt/a(t)$ that increases from 0 as we move into the past. We also define a dimensionless version $\hat{\eta} = \eta H_0$. In our code, we determine the evolution of $\hat{\eta}(a)$, $\phi(a)$, and $v(a)$ via first order equations
\begin{equation}
\label{ODEs}
\begin{split}
{d \phi \over d a} &=-{ v \over a \hat{H}} \\
{d v \over da} &= {V'(\phi) \over a \hat{H}} - 3 {v \over a} \qquad \qquad \hat{H} = \sqrt{{\Omega_M \over a^3} + {1 \over 2} v^2 + V(\phi)} \\
{d \hat{\eta} \over da} &= -{1 \over a^2 \hat{H}} 
\end{split}
\end{equation}
with initial conditions 
\begin{equation}
\phi(1) = 0 \qquad v(1) =  \sqrt{2(1 - \Omega_M - V_0)} \qquad \eta(1) = 0 \; .
\end{equation}

\subsection{Observational data}

In this work, we make use of the most up-to-date catalogues of supernova and BAO observations currently available to test various cosmological models and constrain their parameters. 
The first of these is the DES-SN5YR supernova dataset produced by the Dark Energy Survey (DES) Supernova Program \cite{DES:2024upw, DES:2024hip, DES:2024jxu}.\footnote{This data can be accessed at \href{https://github.com/des-science/DES-SN5YR}{https://github.com/des-science/DES-SN5YR}.} Compared to the earlier Pantheon+ compilation \cite{Scolnic:2021amr, brout2022pantheon+} (used in our previous analysis \cite{VanRaamsdonk:2023ion}), the DES-SN5YR catalogue offers notable advantages in terms of homogeneity, statistical power, and redshift distribution. 
In detail, the Pantheon+ data set is constituted by 1550 unique, spectroscopically confirmed type Ia supernovae spanning a broad redshift range (from very low redshift SNe out to $z \sim 2.3$), though its composition draws from 18 different surveys, making a well-controlled cross-survey calibration apparatus essential. 
By contrast, DES-SN5YR comprises a uniformly analyzed sample of 1635 high-redshift supernovae ($0.1 < z < 1.13$) from the Dark Energy Survey, the largest such collection from a single telescope to date, supplemented by a rigorously selected and recalibrated subset of 194 low-redshift SNe. This results in a particularly dense and coherent sampling of the intermediate redshift regime ($z \gtrsim 0.5$), with a five-fold increase in sample size in this range compared to the Pantheon+ data set, a feature which is of particular interest for probing dynamical dark energy. 
Despite relying on photometric rather than spectroscopic classification, the DES-SN5YR analysis controls classification uncertainties statistically, and the resulting cosmology analysis in \cite{DES:2024hip, DES:2024jxu} is dominated by statistical rather than systematic uncertainties. The constraining power of the DES-SN5YR data is in evidence in the improved significance for dynamical dark energy in the recent DESI analysis \cite{DESI:2024mwx, DESI:2025zgx} combining DES-SN5YR with BAO+CMB observation, as compared to a comparable analysis using Pantheon+ data. 

Our study also utilizes the second data release (DR2) from the Dark Energy Spectroscopic Instrument (DESI) \cite{DESI:2025zpo, DESI:2025zgx},\footnote{This data was accessed at \href{https://github.com/CobayaSampler/bao_data}{https://github.com/CobayaSampler/bao\_data}.} which includes over 14 million high-quality redshifts from galaxies and quasars, along with Lyman-$\alpha$ forest absorption spectra from more than 820,000 quasars, all collected during the first three years of DESI’s main survey operations. This dataset significantly improves upon previous spectroscopic BAO samples, including those of the Sloan Digital Sky Survey (SDSS) used in our previous cosmology analysis \cite{VanRaamsdonk:2023ion}. 
For comparison, our earlier work drew on data produced by the Baryon Oscillation Spectroscopic Survey (BOSS) of SDSS-III and the extended eBOSS of SDSS-IV, which together provide spectroscopic measurements of around 2 million galaxies with redshift coverage $z \lesssim 1.1$, and around 350,000 quasars in range $0.8 < z < 2.2$. DESI DR2 significantly increases upon these statistics and unifies them within a single, coherently designed and executed survey. It maintains a broad redshift range $0.1 \lesssim z \lesssim 3.5$, 
providing in particular more uniform and dense coverage at high redshifts, especially for quasars and Lyman-$\alpha$ forest studies, compared to SDSS and DESI DR1.

\subsection{Constraining parameters}

In this section, we review the methods used to constrain parameters. We use the same methodology as in standard cosmology analysis e.g. in \cite{DESI:2025zgx} and \cite{descollaboration2024darkenergysurveycosmology} and check our methods by reproducing parameter constraints from those previous studies.

\subsubsection{Supernova data versus theory}

For each supernova, we have a redshift value $z$ which has been corrected for the effects of velocities of objects relative to the cosmological background. The redshift is directly related to the scale factor at the time of the supernova by
\begin{equation}
\label{ztoa}
    a = \frac{1} {1 + z} \: .
\end{equation}

The data also provides a ``distance modulus'' $\mu$ related to 
the apparent magnitude $m_B$ of the supernova relative to its absolute magnitude $M$,
\be
\mu = m_B - M \; .
\ee
In the data set, a fiducial value for $M$ is used when calculating $\mu$, but $M$ is treated as a free parameter in the analysis by adding a correction 
\be
(\mu_{\text{data}})_i = \mu^{0}_i - \delta M \; .
\ee

To obtain the theory prediction for $\mu$ given $z$, we use that the distance modulus parameter $\mu$ is related to the inferred luminosity distance $d_L$ via
\begin{equation}
\mu = 5 \log_{10}\left( \frac{d_L}{10 \: \textnormal{pc}} \right) \; ,
\end{equation}
where pc is parsec. The luminosity distance appearing in the distance modulus is in turn related to the (positive) conformal time $\eta$ of the supernova relative to the present time via
\begin{equation}
\label{dtoneta}
      \frac{d_{L}}{1 + z} = \eta = \frac{\hat{\eta}}{H_0} \: , 
\end{equation}
where $\eta$ on the left is the dimensionless quantity appearing in the equations \eqref{ODEs}. Finally, $\hat{\eta}$ can be determined from $a$ and the model parameters via the equations \eqref{ODEs}. Thus we calculate $z \to a \to \eta \to d_L \to \mu$ via the equations above in order to determine the theory prediction for $\mu$. Putting this all together, we get\footnote{The DES data release includes ``Hubble diagram'' redshifts $z_{\text{HD}}$ and ``heliocentric'' redshifts $z_{\text{Hel}}$. Following \cite{DES:2024jxu}, we use $(1 + z_{\text{Hel}}) \hat{\eta}(z_{\text{HD}})$ in our calculations of $\mu$.}
\be
(\mu_{\text{theory}})_i = 5 \log_{10} \left({(1 + z_i) \over H_0 \cdot 10 \: {\rm pc} } \hat{\eta}\left[{1 \over 1 + z_i}\right]\right) \; .
\ee

We note that when taking the difference $\mu_{\text{theory}} - \mu_{\text{data}}$, the parameter $\delta M$ and the parameter $H_0$ only appear in the combination $\delta M - 5 \log_{10} H_0$, so for the analysis, we can set $H_0$ to some fiducial value. In particular, the analysis here will not constrain the actual value of $H_0$.

\subsubsection{BAO data versus theory}

Each BAO data point provides values for an effective redshift $z_{\text{eff}}$ and one of three inferred measures of distance, each of which is expressed as a multiple of $r_{\text{d}}$, the sound horizon at the drag epoch. Theoretical expectations for the provided distance measures can be obtained in terms of the dimensionless variables $\hat{H}$ and $\hat{\eta}$ appearing in equations (\ref{ODEs}) as
\begin{equation}
\begin{split}
    D_H \equiv {d_H \over r_d} &= {c \over H_0 r_d} {1 \over \hat{H}} \\
    D_M \equiv  {d_M \over r_d} &= {c \over H_0 r_d} \hat{\eta} \\
    D_V  \equiv {d_V \over r_d} &= {c \over H_0 r_d} \left({z \hat{\eta}^2 \over \hat{H}} \right)^{1 \over 3}
\end{split}
\end{equation}
where $d_V$ is a volume-averaged angular diameter distance. For comparison with theory, we compute $\hat{H}$ and $\hat{\eta}$ for the given value of $z$ via the equations (\ref{ODEs}) and use these on the right hand side of the previous equations to find the theory values for $D_H$, $D_M$, and $D_V$. 

The right hand sides above depend only on the combination $H_0 r_d$. Thus, we can take a fiducial value for $H_0$ (as above) and treat $r_d$ as a free parameter in the analysis.

\subsubsection{Parameter estimation}

To evaluate the fit of a certain set of model parameters to the supernova and BAO data, we calculate a $\chi^2$ value
as
\begin{equation}
\chi^2 = \chi^2_{\text{SN}} + \chi^2_{\text{BAO}} \; .
\end{equation}
Here
\be
\label{chisquaredS}
\chi^2_{\text{SN}} = (\mu_{\text{data}} - \mu_{\text{model}})_i (C_{\text{SN}})^{-1}_{ij} (\mu_{\text{data}} - \mu_{\text{model}})_j \: ,
\ee
where $C_{\text{SN}}$ is the covariance matrix provided by the DES-SN5YR results that takes into account statistical and systematic error. For the BAO contribution, we have
\be
\label{chisquaredB}
\chi^2_{\text{BAO}} = (D - D_{\text{model}})_i (C_{\text{BAO}})^{-1}_{ij} (D - D_{\text{model}})_j \: ,
\ee
where the various $D$ values in the DESI DR2 data set are either $D_V$, $D_H$, or $D_M$, and $C_{\text{BAO}}$ is the covariance matrix provided by DESI DR2.

According to the standard analysis, the $\chi^2$ is used to assign a likelihood
\be
\label{likelihood}
{\cal L} = e^{- \chi^2 /2}
\ee
to a set of model parameters, where we assume a flat prior distribution of parameters apart from the constraints $\Omega_M > 0$ and $V_0 + \Omega_M < 1$, the latter required by positivity of scalar kinetic energy. 

Since $\chi^2_{\text{SN}}$ is quadratic in the parameter $\delta M$, we can analytically marginalize over this parameter to obtain a likelihood distribution for the remaining parameters \cite{Goliath:2001af}. Explicitly, we have 
\begin{equation}
\begin{split}
e^{- (\chi_{\text{SN}}^{2})_{\text{eff}} /2} & = \int d (\delta M) \: e^{- \left[ \chi_{\text{SN}}^2(\delta M)\right] /2} \\
& = \int d (\delta M) \: e^{-  (\mu^0 - \delta M v -  \mu_{\text{model}})_i (C_{\text{SN}})^{-1}_{ij} (\mu^0- \delta M v - \mu_{\text{model}})_i}
\end{split}
\end{equation}
where $v_i = 1$ for all $i$. Integrating gives
\begin{multline}
(\chi_{\text{SN}}^{2})_{\text{eff}} = (\mu^0 - \mu_{\text{model}})_i (C_{\text{SN}})^{-1}_{ij} (\mu^0 - \mu_{\text{model}})_j  \\
-  {|v_i (C_{\text{SN}})^{-1}_{ij} (\mu^0 - \mu_{\text{model}})_j|^2 \over v_i (C_{\text{SN}})^{-1}_{ij} v_j} + \ln \left({v_i (C_{\text{SN}})^{-1}_{ij} v_j \over 2 \pi} \right) \; .
\end{multline}

In our analysis, we use a Markov Chain Monte Carlo (MCMC) algorithm to sample from the probability distribution (\ref{likelihood}) calculated using $(\chi_{\text{SN}}^{2})_{\text{eff}}$, $\chi_{\text{BAO}}^2$, or the combination  $(\chi_{\text{SN}}^{2})_{\text{eff}} + \chi_{\text{BAO}}^2$. The parameters are $(\Omega_M,V_0,V_1)$ for the supernova-only constraints, and $(\Omega_M,V_0,V_1, r_{\text{d}})$ for the BAO and combined constraints. The Monte Carlo simulation yields an ensemble of parameter values that should follow the $\exp(-\chi^2/2)$ distribution. From this, we can calculate mean values of the parameters and the associated standard deviations, or produce histograms and correlation plots.

\section{Results} \label{sec:results}

In this section, we present the results of our analysis, initially taking a linear approximation $V(\phi) = V_0 + V_1 \phi$ to the potential but later investigating the effects of including a quadratic term.

\subsection{Linear potential constraints}

\begin{figure}
    \centering
    \includegraphics[width= 0.9 \textwidth]{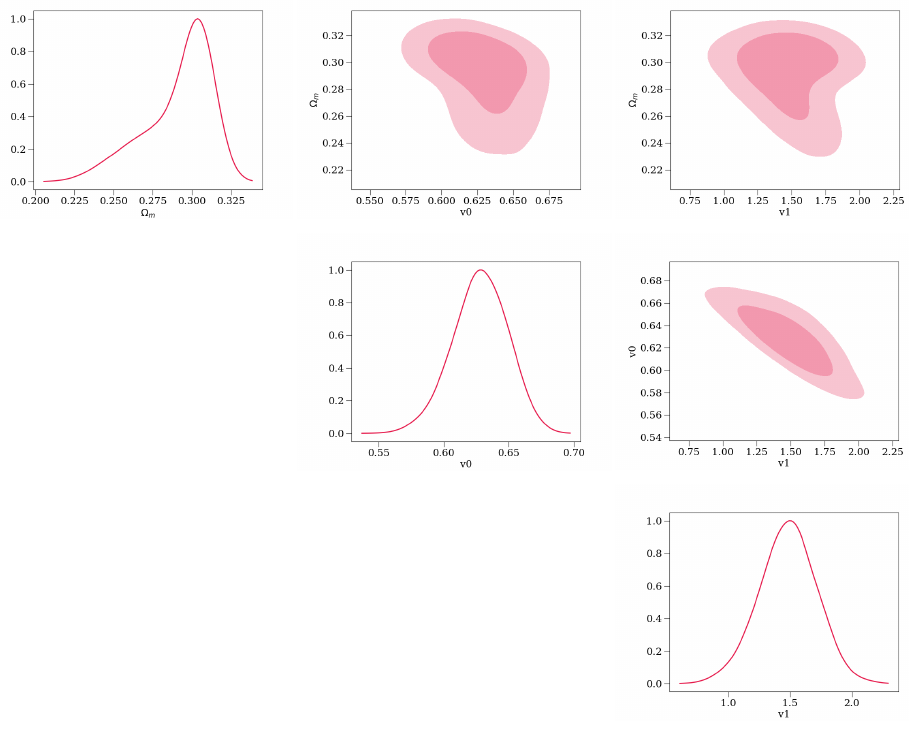}
    \caption{Constraints on parameters $\Omega_M$, $V_0$, and $V_1$ for a linear potential scalar field model, using $e^{-\chi^2 / 2}$ likelihood calculated from combined DES-SN5YR supernova and  DESI DR2 BAO data. The $\Lambda$CDM model corresponds to $V_1 = 0$, $\Omega_M + V_0 = 1$. }
    \label{fig:triangle}
\end{figure}

The results for the parameter constraints for the linear potential model are summarized in Table \ref{tab:linmodel} and Figure \ref{fig:triangle}.

\begin{table}[h]
\centering
\begin{tabular}{lccc}
\hline
\textbf{Data} & $\Omega_M$ & $V_0$ & $V_1$ \\
\hline
DES-SN5YR SN     & $0.253 \pm 0.050$ & $0.518 \pm 0.073$ & $3.8 \pm 1.0$ \\
DESI DR2 BAO   & $0.299 \pm 0.048$ & $0.442 \pm 0.156$ & $2.8 \pm 1.0$ \\
BAO + SN      & $0.292 \pm 0.021$ & $0627 \pm 0.022$ & $1.489 \pm 0.250$ \\
\hline
\end{tabular}
\caption{Linear potential model parameter constraints based on supernova data only, BAO data only, and combined SN and BAO data sets.}\label{tab:linmodel}
\end{table}

\begin{figure}
    \centering
    \includegraphics[width= 1 \textwidth]{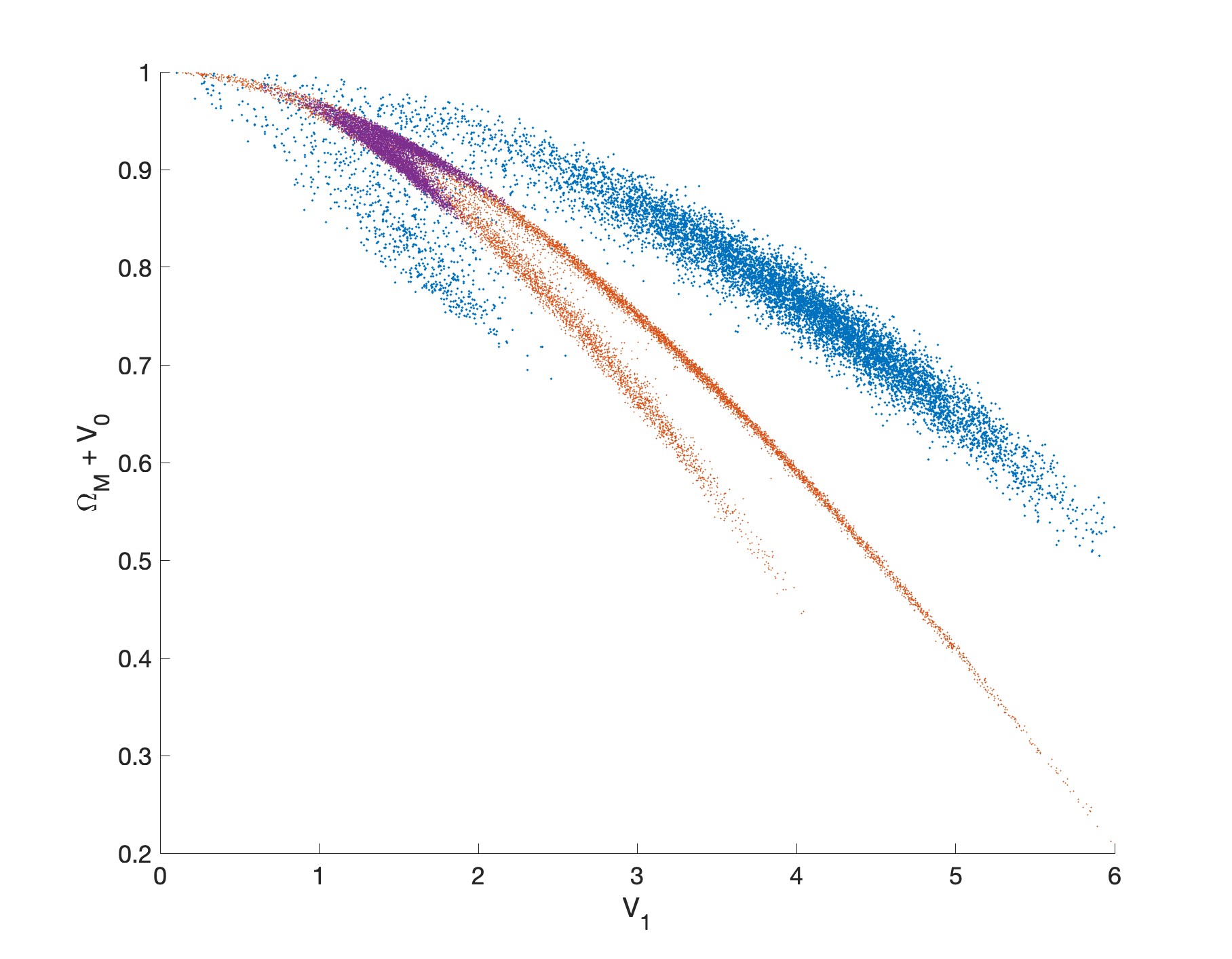}
    \caption{Distribution of parameter values $[\Omega_M + V_0,V_1]$ for a linear potential scalar field model, using $e^{-\chi^2 / 2}$ likelihood calculated from DES-SN5YR supernova data (blue), DESI DR2 BAO data (orange), and combined data (purple). Each distribution includes 9,000 points. The $\Lambda$CDM model corresponds to $V_0 + \Omega_M = 1$, $V_1 = 0$. The two-pronged structure of the distribution is explained in the text.}
    \label{fig:combnice}
\end{figure}

It may be initially surprising that the $V_1$ range for the combined dataset lies significantly below that from both the SN and the BAO data sets individually. However, Figure \ref{fig:combnice}, showing the distribution of values for $[\Omega_M + V_0, V_1]$ provides insight here. The figure indicates that for both BAO data and SN data, we have an approximate degeneracy, with a relatively wide range of possible $V_1$ values but $\Omega_M + V_0$ values strongly constrained by the choice of $V_1$. Further, both the SN distribution and the BAO distribution show two distinct branches, with smaller and larger values of $V_1$ for the same value of $\Omega_M + V_0$. This has a simple explanation. Models in the larger $V_1$ branch have a scalar field that initially moves upward on the potential before falling back down, while models in the smaller $V_1$ branch have a monotonic behavior for the scalar field where it simply descends the potential. The lower value for $V_1$ in the combined analysis follows from the fact that while the distributions for both SN and BAO have a large weight at higher $V_1$ values, the distributions only intersect at smaller $V_1.$ 

\begin{figure}
    \centering
    \includegraphics[width= 1 \textwidth]{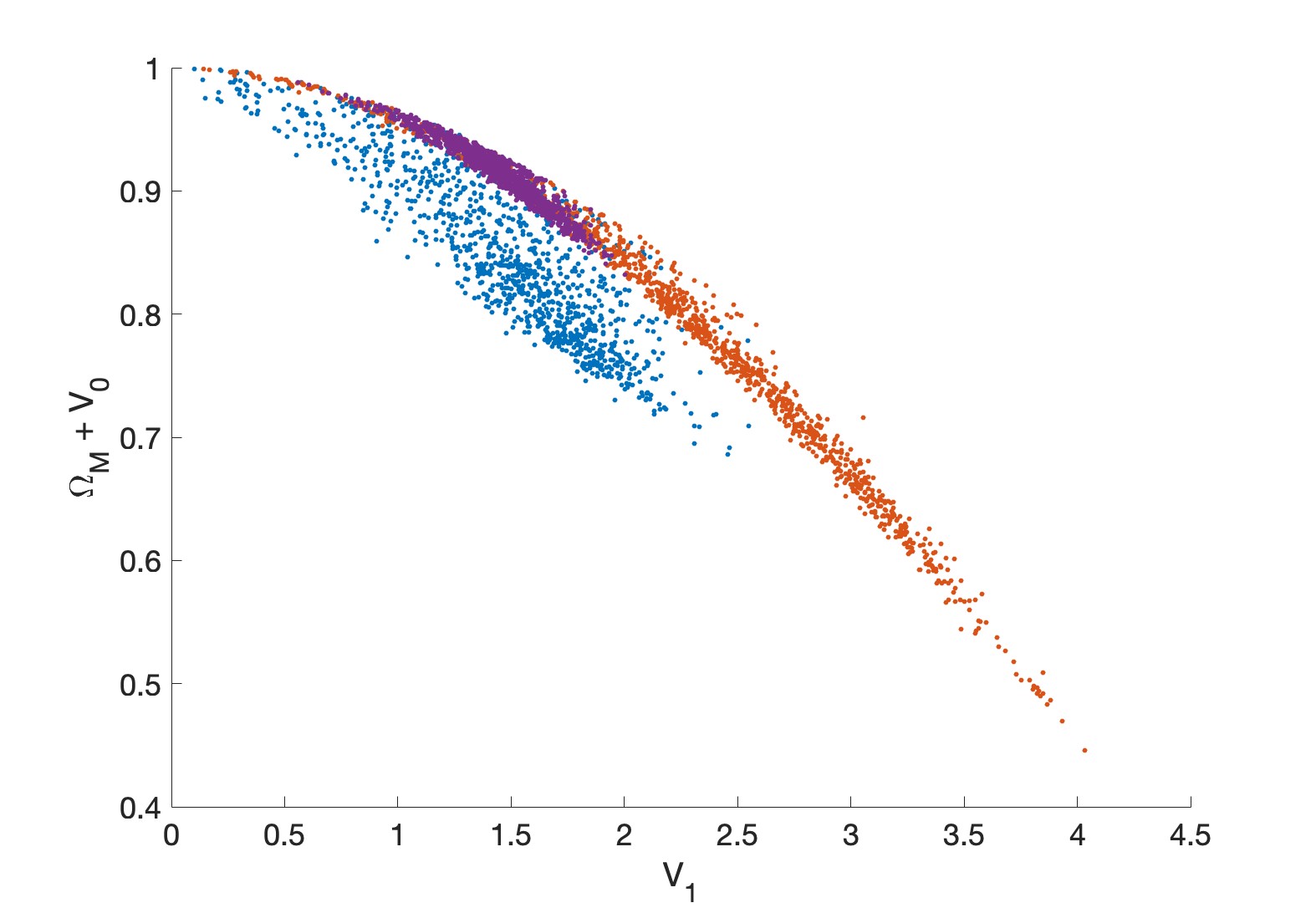}
    \caption{Distribution of parameter values $[\Omega_M + V_0,V_1]$ for a linear potential scalar field model with a prior that $\phi(t)$ is monotonic. The points follow an $e^{-\chi^2 / 2}$ distribution calculated from DES-SN5YR supernova data (blue), DESI DR2 BAO data (orange), and combined data (purple). Each distribution includes 1,100 points. The $\Lambda$CDM model corresponds to $V_0 + \Omega_M = 1$, $V_1 = 0$.}
    \label{fig:combpos}
\end{figure}

We can also look at the results using the prior that only models with a monotonically descending scalar field be considered. The class would be expected under the assumption that the scalar started at some high value of its potential in the early universe and has been descending ever since. Figure \ref{fig:combpos} is the version of Figure \ref{fig:combnice} with this prior.

The parameter constraints with the scalar monotonicity prior are summarized in Table \ref{tab:mono_prior} below. We see that the results for the combined datasets are very similar to those without the prior.
\begin{table}[h]
\centering
\begin{tabular}{lccc}
\hline
\textbf{Data} & $\Omega_M$ & $V_0$ & $V_1$ \\
\hline
DES-SN5YR SN     & $0.144 \pm 0.097$ & $0.706 \pm 0.045$ & $1.47 \pm 0.44$ \\
DESI DR2 BAO   & $0.254 \pm 0.041$ & $0.532\pm 0.089$ & $2.27 \pm 0.75$ \\
BAO + SN     & $0.282 \pm 0.022$ & $0.635 \pm 0.019$ & $1.42 \pm 0.22$ \\
\hline
\end{tabular}
\caption{Linear potential model parameter constraints  with a prior restricting to models with a monotonic scalar field evolution.}
\label{tab:mono_prior}
\end{table}

\begin{figure}
    \centering
    \includegraphics[width= 0.6 \textwidth]{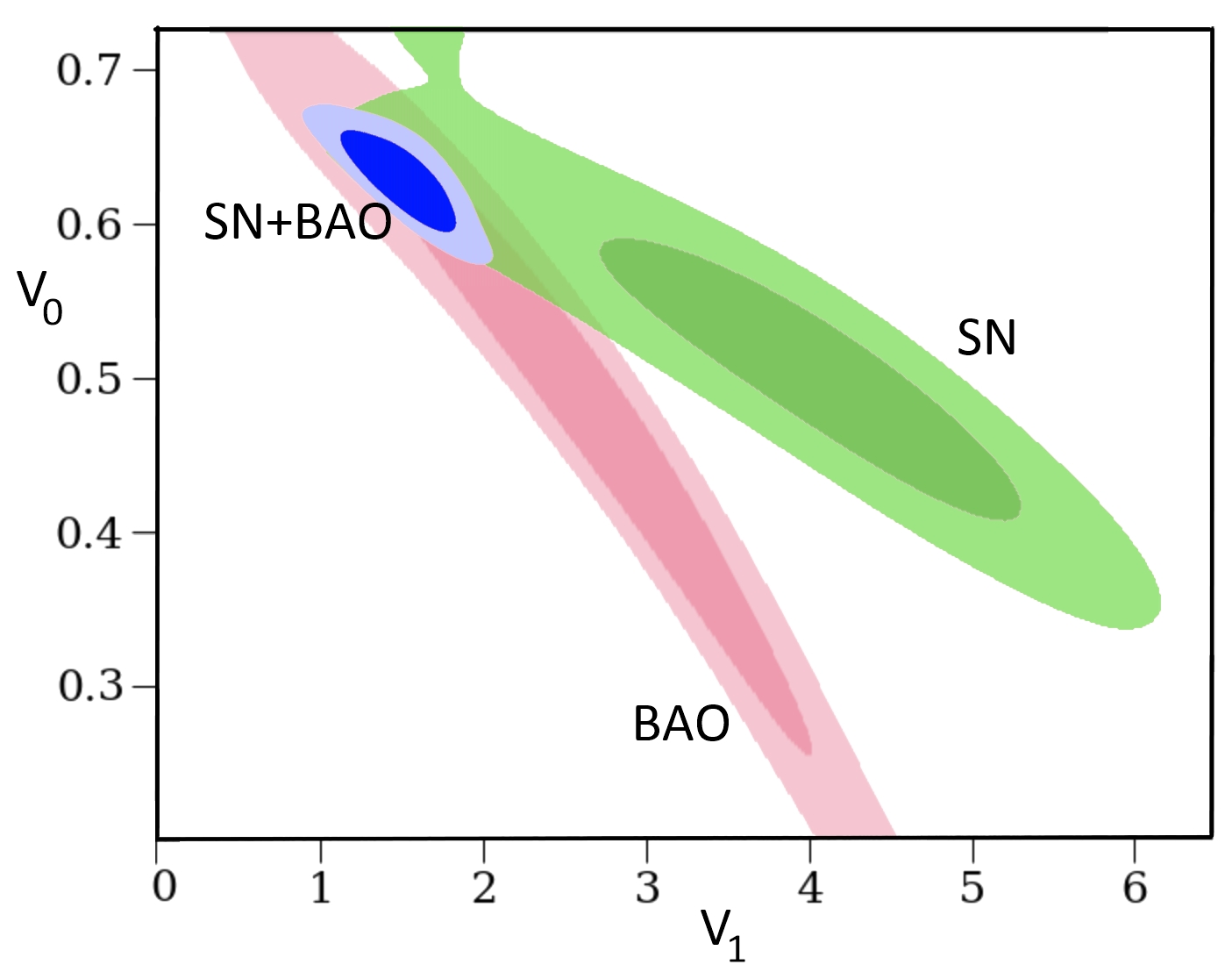}
    \caption{Distribution of potential parameter values $V_0$ and $V_1$ for a linear potential scalar field model constrained by DES-SN5YR supernova data (green), DESI DR2 BAO data (pink), and combined data (blue). For each dataset, we show equal $\chi^2$ contours containing 68\% and 95\% of the $\exp(-\chi^2/2)$ distribution.}
    \label{fig:combplot1}
\end{figure}

Focusing now on the potential parameters, Figure \ref{fig:combplot1} shows contours of equal density in the distribution marginalized over the other parameters containing 68\% and 95\% of the distribution. Particularly striking is that, using the combined dataset, these contours lie well away from the value $V_1 = 0$ that we have in the $\Lambda$CDM model. This was also reflected in parameter constraint $V_1 = 1.49 \pm 0.25$ or $V_1 = 1.42 \pm 0.22$ with the scalar monotonicity prior. In either case, the value $V_1 = 0$ is more than five standard deviations from the mean. This would seem to strongly disfavour $\Lambda$CDM in the context of the linear potential models. The difference between $\chi^2$ in the best fit $\Lambda$CDM model and the best fit linear potential model is 
\begin{equation}
\Delta \chi^2 \equiv \chi^2_{V_0V_1} - \chi^2_{\Lambda\text{CDM}} \approx -13.7 \: ,
\end{equation}
corresponding to $3 \sigma$ preference for the linear potential model in an approach that does not take into account parameter space volumes.  

\begin{figure}
    \centering
    \includegraphics[width= 0.6 \textwidth]{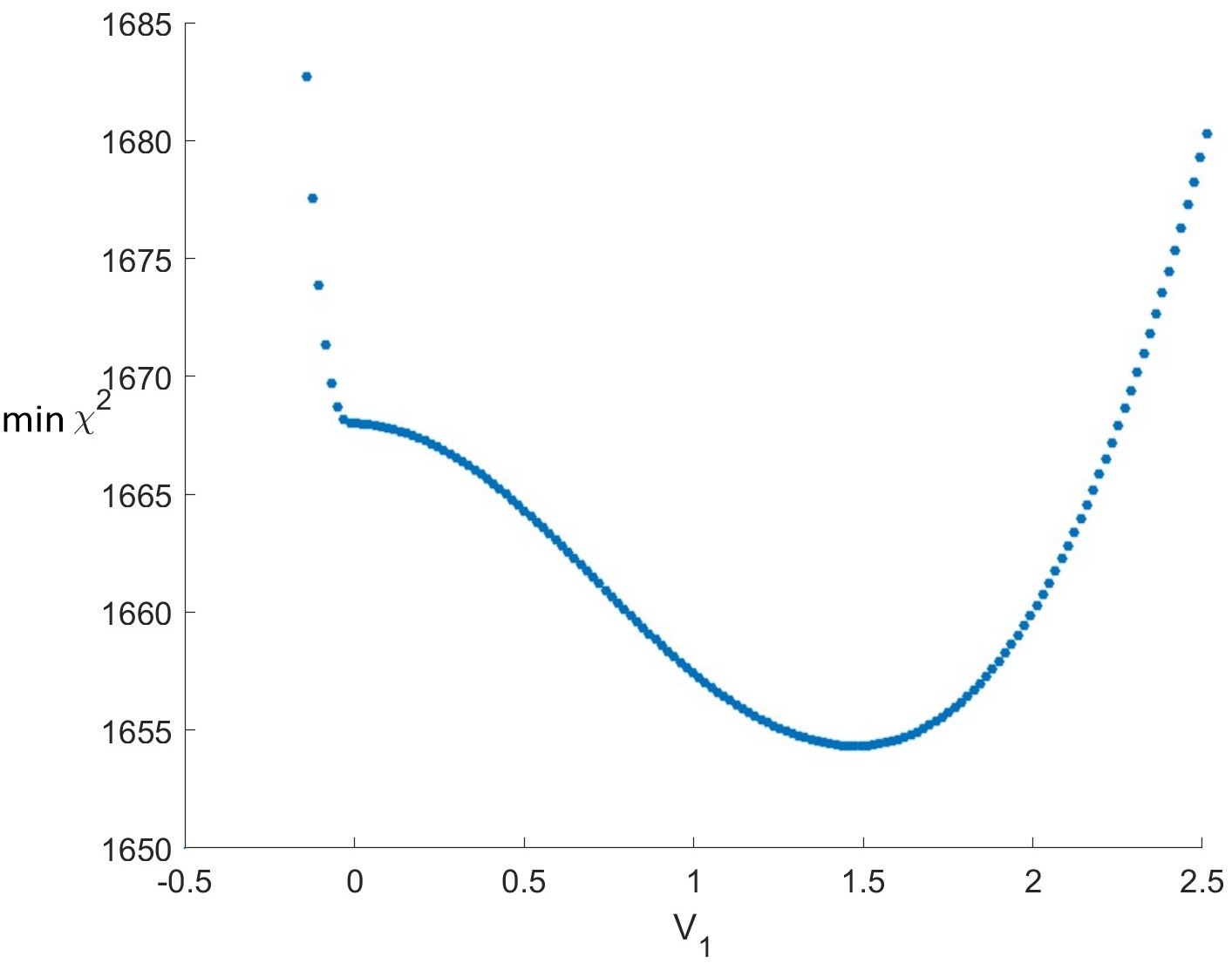}
    \caption{Value of $\chi^2$ for the best fit linear potential model with various values of $V_1$. Here, we still allow for scalar kinetic energy, but the best fit $\Lambda$CDM model gives a similar $\chi^2$ to the best fit $V_1=0$ model overall.}
    \label{fig:V1chi}
\end{figure}

The apparent difference in significance levels can be understood via Figure \ref{fig:V1chi}, where we plot $\chi^2$ for the best fit model with various values of $V_1$. We see that $\chi^2$ increases more slowly as $V_1 \to 0^{+}$ than the quadratic approximation to the function around its minimum, reflecting non-Gaussianity in the distribution. 

A notable aspect of figure \ref{fig:V1chi} is the rapid increase in $\chi^2$ for $V_1 < 0$, i.e. the case where the potential is increasing at the present time. This is understandable since in order for the scalar to be ascending its potential now, it would have had to be at significantly lower values in the past and would have had substantially more kinetic energy, since it is losing kinetic energy both to potential energy and to the Hubble damping as it rises upwards. This leads to a very different qualitative behavior for the evolution. In section \ref{sec:mock} below, we see that this effect gives rise to a bias towards positive $V_1$ in the parameter estimation, though not nearly enough to explain our results.

The results in the present analysis favor a linear potential model over $\Lambda$CDM substantially more than in our previous analysis \cite{VanRaamsdonk:2023ion} using the Pantheon+ supernova database and a collection of older BAO data. In that case, the result $V_1 = 1.1 \pm 0.3$ corresponded to roughly $3 \sigma$ significance according to the Bayesian approach, but we we had only $\Delta \chi^2 = -3.7$.

\subsection{Analyzing mock data and positive $V_1$ bias}
\label{sec:mock}

To further assess the significance of our results for $V_1$ and to test for potential bias in the parameter estimation procedure, we repeated our analysis on a series of fabricated data sets generated by starting with a $\Lambda$CDM model (with $\Omega_M = 0.31$, $V_0 = 0.69$, $V_1 = 0$), calculating the theoretical values for $D_i$ and $\mu_i$ using the $z_i$ values in the real BAO and SN data, and then adding Gaussian noise with a covariance structure that reproduces both the variances and correlations in the BAO and SN covariance matrices. 

For each member of an ensemble of 144 fabricated datasets based on the same $\Lambda$CDM model, we performed a full likelihood analysis based on the $V_0 V_1$ model. The parameter estimates and 68\% intervals that we obtained on average were
\[
\Omega_M = 0.297 \pm 0.014 \qquad V_0 = 0.678 \pm 0.016 \qquad V_1 = 0.730 \pm 0.284 \; .
\]
It is notable that the analysis gives a $V_1$ value that is around $2.5 \sigma$ away from the ``true'' value $V_1 = 0$. This can be understood by a combination of factors. First, allowing non-zero values of $V_1$ increases the number of parameters in the model and allows some reduction of $\chi^2$ simply due to overfitting of the random fluctuations introduced into the data. On the other hand, negative values of $V_1$ dramatically change the behavior of the model, as seen in Figure \ref{fig:V1chi}, so the distribution of parameter values able to adequately fit the data is expected to have support mostly for positive $V_1$. Thus, we end up with a biased distribution of $V_1$ values. 

In order to properly assess the significance of our result $V_1 = 1.49$ for the real data in light of the positive $V_1$ bias, we can compare our measured value to the distribution of $V_1$ mean values obtained in ensemble of mock data sets. This distribution has a mean of 0.73 and a standard deviation of 0.19. Thus, our measured value of $V_1 = 1.49$ for real data lies $4 \sigma$ above the mean in the distribution of $V_1$ values produced by mock $\Lambda$CDM data. According to this, the likelihood that our measured value is arising by chance from $\Lambda$CDM data with errors having the same covariance as the real data is only 0.01\%. 

To verify that our parameter estimation procedure would provide correct results in the case that the data actually arises from a linear potential model, we also created fabricated data sets based on a model universe governed by a linear potential model with our estimated values of $\Omega_M = 0.292$, $V_0 = 0.627$, $V_1 = 1.49$ (again introducing noise consistent with the covariance matrices). Performing our likelihood analysis on this data, we found average parameter estimates
\[
\Omega_M = 0.274 \pm 0.022 \qquad V_0 = 0.620 \pm 0.023 \qquad V_1 = 1.71 \pm 0.23 \; ,
\]
so while the procedure slightly overestimates $V_1$, the estimated $1 \sigma$ parameter ranges include the ``true'' values in each case.

\subsection{Comparison with $w_0w_a$}

It is interesting to compare our results with the parameter constraints in the frequently used $w_0w_a$ model. Constraining parameters for this model with the same combined dataset, we find $w_0 = -0.781 \pm 0.069$ and $w_a = -0.728 \pm 0.449$, in good agreement with the results ( $w_0 = -0.781^{+ 0.067}_{- 0.076}$ and $w_a = -0.72 \pm 0.47$) presented in \cite{DESI:2025zgx} for the same combination of data. This provides a good check of our procedures.

\begin{figure}
    \centering
    \includegraphics[width= 1 \textwidth]{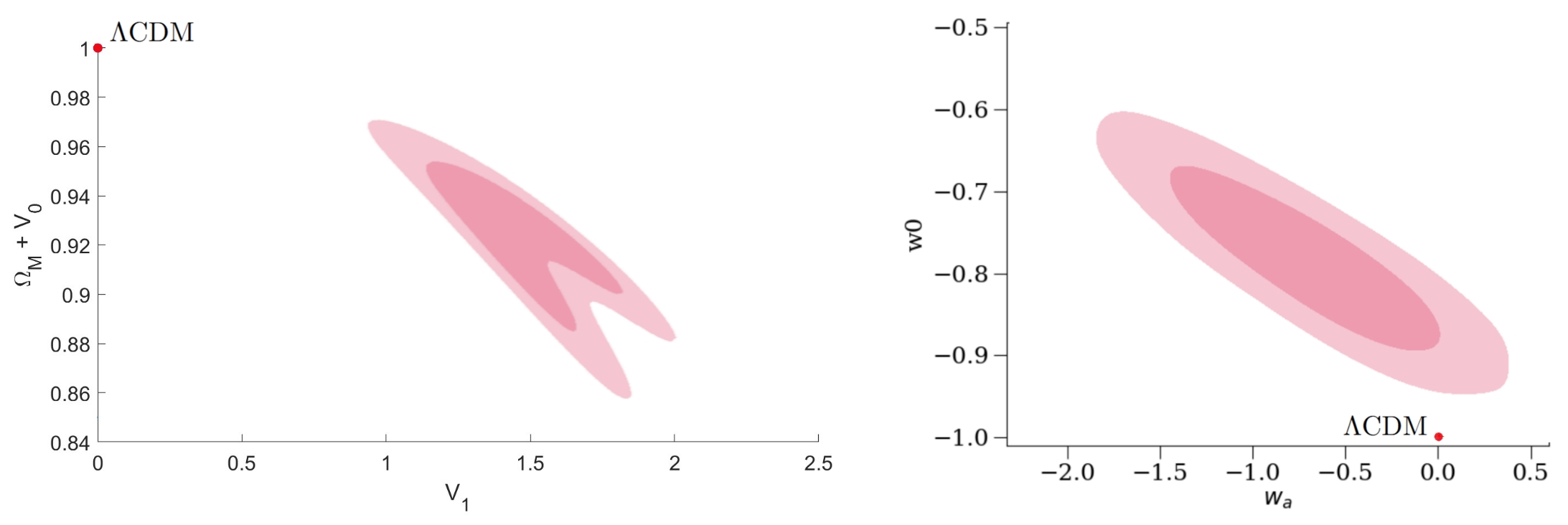}
    \caption{Comparison of 68\% and 97\% likelihood contours for $V_{0}V_{1}$ (linear potential) and $w_{0}w_{a}$CDM extensions of $\Lambda$CDM. The $\Lambda$CDM model is indicated by a red point in each case.}
    \label{fig:V0V1w0wa}
\end{figure}

We find that the linear potential models provide a slightly better fit to the data compared with the best fit $w_0w_a$ models, which have $\Delta \chi^2 = -13.5$ compared with $\Lambda$CDM. On the other hand, the $\Lambda$CDM model lies much further on the outskirts of the $\exp(-\chi^2/2)$ distribution for the linear potential model, as seen in Figure \ref{fig:V0V1w0wa}.

\subsection{Quadratic model}

\begin{figure}
    \centering
    \includegraphics[width= 0.6 \textwidth]{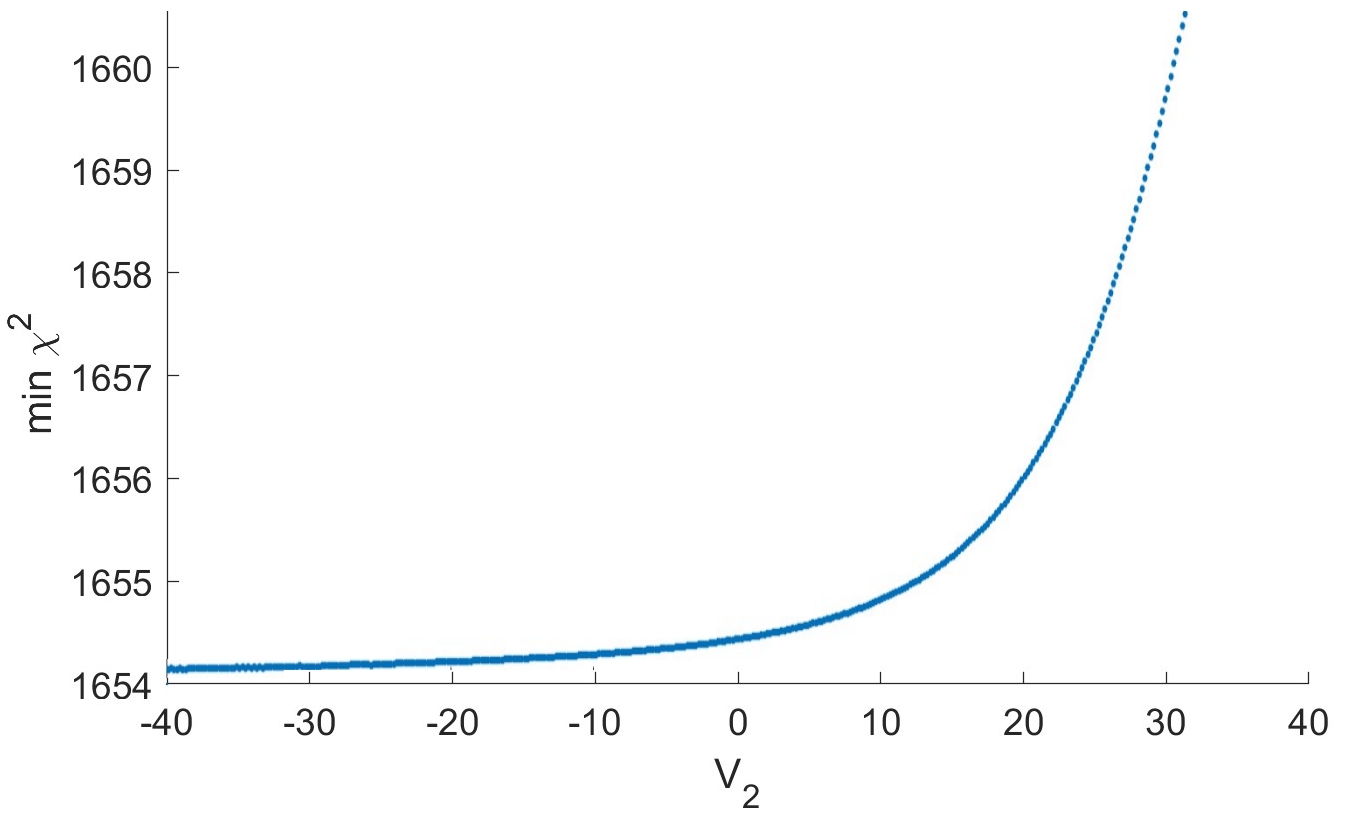}
    \caption{Value of $\chi^2$ for the best fit quadratic potential model with various values of $V_2$.}
    \label{fig:V2chi}
\end{figure}

We have also investigated the effects of extending the model to include a quadratic term in the scalar potential
\begin{equation}
V(\phi) = V_0 + V_1 \phi + {1 \over 2} V_2 \phi^2 \: .
\end{equation}
As far as we could tell, adding the extra parameter does not allow a significant decrease in $\chi^2$ compared with the linear potential model, with $\Delta \chi^2 < 1$ compared to the best fit linear potential model for all models tested. Figure \ref{fig:V2chi} shows the minimum $\chi^2$ as a function of $V_2$. 

Here, we have focused our investigation on models that are close to the best fit linear potential models. In \cite{VanRaamsdonk:2023ion}, we also found models with significantly larger values of $|V_{2}|$ that provide a good fit to data. For large positive $V_{2}$, we found models 
where a scalar field oscillating about the minimum of its potential gives a contribution to the energy that replaces some of the matter. For large negative $V_{2}$, we found finely tuned models where the scalar sits near the top of the negative quadratic potential before sliding downward at late times. It would be interesting to investigate whether models of these types can provide a good fit to the data that we analyze in this work.          

\subsection{Results from reduced data sets} \label{sec:red_data_sets}

\begin{figure}
    \centering
    \includegraphics[width= 0.6 \textwidth]{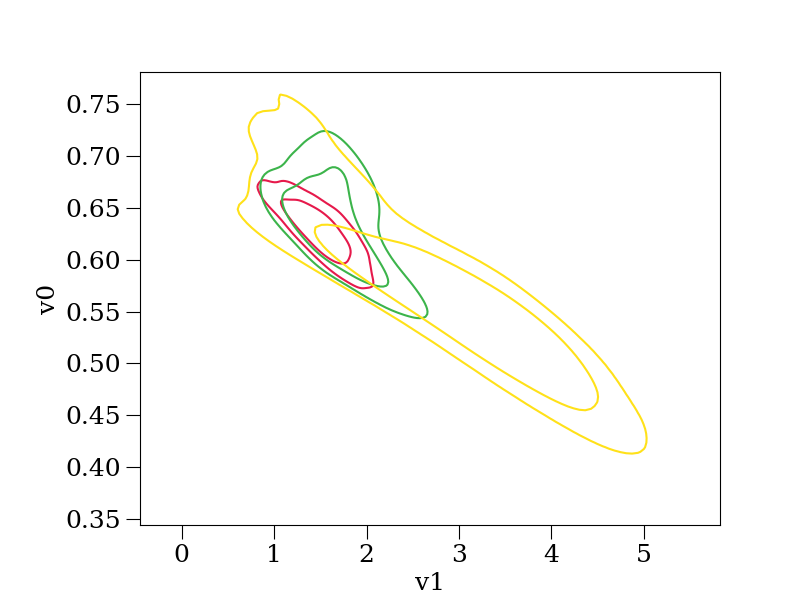}
    \caption{Contour plots of 68\% and 95\% confidence regions for parameters $V_{0}, V_{1}$ in linear potential model, using BAO + SN data with no redshift cut (red), high-redshift cut $z < 1$ (green), and high-redshift cut $z < 0.8$ (yellow).}
    \label{fig:highzcut}
\end{figure}

\begin{figure}
    \centering
    \includegraphics[width= 0.6 \textwidth]{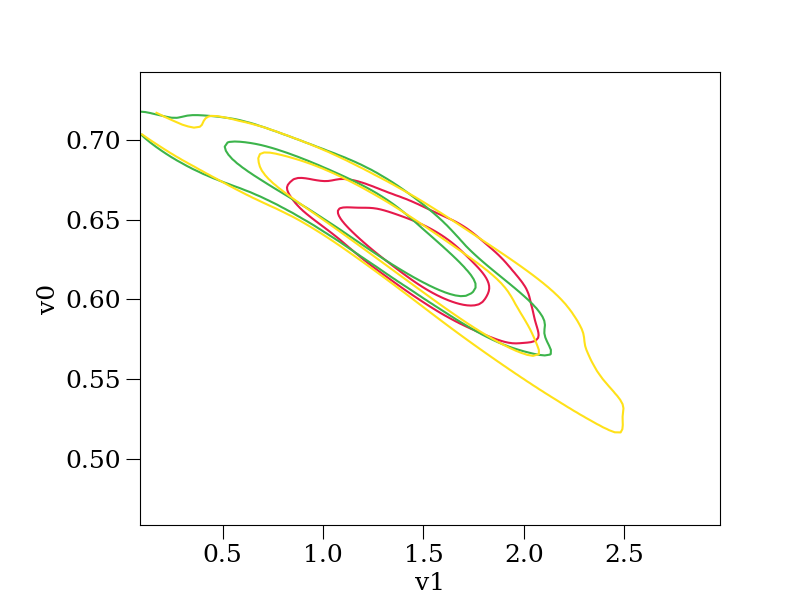}
    \caption{Contour plots of 68\% and 95\% confidence regions for parameters $V_{0}, V_{1}$ in linear potential model, using BAO + SN data with no redshift cut (red), low-redshift cut $z > 0.1$ (green), and low-redshift cut $z > 0.2$ (yellow). The ``clipping'' on the left of the green and yellow 95\% contours is a result of the prior $\Omega_{M} + V_{0} < 1$.}
    \label{fig:lowzcut}
\end{figure}

To better understand how the linear potential model improves the fit to data as compared with $\Lambda$CDM, and also as a further check on the robustness of the results, we repeated our 
analysis with various cuts to the range of redshifts being used. Eliminating larger redshift data, we find results for the 68\% and 95\% contours shown in Figure \ref{fig:highzcut}, and the following parameter constraints:
\begin{table}[h]
\centering
\begin{tabular}{lccc}
\hline
\textbf{Cut} & $\Omega_M$ & $V_0$ & $V_1$ \\
\hline
No cut     & $0.290 \pm 0.023$ & $0.628 \pm 0.021$ & $1.49 \pm 0.25$ \\
\hline
$z < 1$     & $0.258 \pm 0.059$ & $0.634 \pm 0.038$ & $1.67 \pm 0.36$ \\
\hline
$z < 0.8$     & $0.284 \pm 0.048$ & $0.565 \pm 0.068$ & $2.87 \pm 1.02$ \\
\hline
\end{tabular}
\caption{Marginalized parameter constraints for linear potential model with BAO + SN data, with various high-redshift cuts applied.}
\label{tab:highzcut}
\end{table}

Apparently, excluding higher redshift data tends to favour even higher values of $V_{1}$ than with the full dataset, though the results are still in reasonable agreement with the results from the full data set. 

If we instead consider removing low $z$ data, we find the 
results in Figure \ref{fig:lowzcut} and Table \ref{tab:lowzcut}.
\begin{table}[h]
\centering
\begin{tabular}{lccc}
\hline
\textbf{Cut} & $\Omega_M$ & $V_0$ & $V_1$ \\
\hline
No cut     & $0.290 \pm 0.023$ & $0.628 \pm 0.021$ & $1.49 \pm 0.25$ \\
\hline
$z > 0.1$     & $0.292 \pm 0.020$ & $0.644 \pm 0.036$ & $1.25 \pm 0.44$ \\
\hline
$z > 0.2$     & $0.290 \pm 0.025$ & $0.630 \pm 0.043$ & $1.41 \pm 0.51$ \\
\hline
\end{tabular}
\caption{Marginalized parameter constraints for linear potential model with BAO + SN data, with various low-redshift cuts applied.}
\label{tab:lowzcut}
\end{table}
Here, we find good agreement for all parameters.

We note that the significance of the results does appear to decrease substantially once the $z < 0.1$ data are removed. With the $z > 0.1$ data, $V_1 = 0$ lies within $3\sigma$ of the mean value, while the $\Delta \chi^2$ between the linear potential model and $\Lambda$CDM is only $\Delta \chi^{2} = - 1.3$. For the $w_0w_a$ model, removing the $z<0.1$ samples gives results $w_0 = -0.89 \pm 0.13$, $w_a = 0.10 \pm 0.78$ that are entirely consistent with the $\Lambda$CDM values $w_0 = -1, w_a = 0$.

It is notable that the $z < 0.1$ samples in the DES-SN5YR dataset are almost all ``historical'' samples that have been added to those found by DES. It is possible that unaccounted for systematic errors in these historical samples could thus be an explanation for the strong significance observed when using the full data set, as previously noted in e.g. \cite{Gialamas:2024lyw, Efstathiou:2024xcq, Colgain:2024mtg,Huang:2025som}.

\section{Discussion} \label{sec:disc}

Assuming that the DES-SN5YR and DESI DR2 data sets provide an accurate picture of the recent background cosmological evolution without unaccounted for systematic errors, our results provide significantly stronger evidence compared with our previous analysis that decreasing dark energy arising from scalar evolution is favored over $\Lambda$CDM. This complements other recent analyses \cite{DESI:2024mwx,DESI:2025zgx} that have found suggestions of decreasing dark energy via the $w_0 w_a$ parameterization. Our $V_0 V_1$ model represents a different two-parameter extension of the $\Lambda$CDM model where $w(a)$ tends to be significantly non-linear in the best fit models and where we have a standard effective field theory description for any values of the parameters. As we have mentioned, the significance relies substantially on the very recent $z < 0.1$ supernova data, so it will be important to evaluate whether there might be some unaccounted-for systematic errors there.

It would be interesting to incorporate additional observations to further constrain the parameters in the $V_0 V_1$ model. We do not have any theoretical reason to believe that this linear form of the potential should remain valid for the full span of cosmological evolution,
so it is most sensible here to focus on data that probes the recent cosmological evolution. One particularly relevant probe is the Integrated Sachs–Wolfe (ISW) effect, which arises from the decay of gravitational potentials at late times and can be measured through cross-correlations between CMB temperature maps and large-scale structure surveys. ISW observations are directly sensitive to the dynamics of dark energy at late times and offer a natural way to test the scalar field evolution described by our model.\footnote{We thank both Douglas Scott and Alex Krolewski for this suggestion.}

From the viewpoint of string theory and/or holographic quantum gravity, the theoretical implications of a definitive demonstration of evolving dark energy would be enormous. It would likely mean that we are much closer to a complete quantum gravity description of nature than if the correct effective theory is $\Lambda$CDM, since evolving dark energy via time-dependent scalars is generic in the class of gravitational theories that we already have a good theoretical description of. Such a description would introduce significant new phenomenological challenges and opportunities, as we have described previously in \cite{Antonini:2022xzo}. A particularly interesting possibility in this case is that the gravitational effective theory is supersymmetric at the negative extremum of the scalar potential, with the Standard Model arising from this supersymmetric model via symmetry breaking due to the scalar field's expectation value and time derivative at the present time. As far as we are aware, such a scenario for phenomenology has not been seriously investigated in the past.

\section*{Acknowledgements}
We would like to thank Douglas Scott and Alex Krolewski for helpful discussions and comments.
MVR acknowledges support from the National Science and Engineering Research Council of Canada
(NSERC) and the Simons Foundation via a Simons Investigator Award. Research
at the Perimeter Institute is supported by the Government of Canada through the Department of Innovation, Science and Industry Canada, and by the Province of Ontario
through the Ministry of Colleges and Universities.

\bibliographystyle{JHEP}
\bibliography{refs}

\end{document}